\documentclass[conference]{IEEEtran}
\IEEEoverridecommandlockouts

\usepackage{amsmath,amsfonts}
\usepackage{algorithmic}
\usepackage{array}
\usepackage[caption=false,font=normalsize,labelfont=sf,textfont=sf]{subfig}
\usepackage{textcomp}
\usepackage{stfloats}
\usepackage{url}
\usepackage{verbatim}
\usepackage{graphicx}
\def\BibTeX{{\rm B\kern-.05em{\sc i\kern-.025em b}\kern-.08em
    T\kern-.1667em\lower.7ex\hbox{E}\kern-.125emX}}
\usepackage{balance}
\usepackage{caption}
\usepackage{siunitx}
\usepackage{tikz}
\usetikzlibrary{shapes.geometric, arrows.meta, positioning, calc, fit}

\usepackage{algorithm}
\usepackage{mathtools}	
\usepackage{amssymb}
\usepackage{amsthm}

\usepackage[utf8]{inputenc}

\usepackage{ifthen}

\usepackage{microtype}

\usepackage{caption}

\usepackage{bm}

\usepackage[draft,nomargin,marginclue,inline]{fixme}
\makeatletter
\renewcommand*\FXLayoutMarginClue[3]{%
  \marginpar[%
  \raggedleft\@fxuseface{margin}\textcolor{red}{\ignorespaces $ \Rightarrow $}]{%
    \raggedright\@fxuseface{margin}\textcolor{red}{\ignorespaces $ \Leftarrow $}}}
\makeatother	

\usepackage{tumcolor}

\usepackage{pgfplotstable}	

\pgfplotsset{
	discard if/.style 2 args={
        x filter/.append code={
            \edef\tempa{\thisrow{#1}}
            \edef\tempb{#2}
            \ifx\tempa\tempb
                
            \fi
        }
    },
    discard if not/.style 2 args={
        x filter/.append code={
            \edef\tempa{\thisrow{#1}}
            \edef\tempb{#2}
            \ifx\tempa\tempb
            \else
                
            \fi
        }
    }
}

\usepackage{cite}

\usepackage[acronym,shortcuts]{glossaries}
\newacronym{cnn}{CNN}{convolutional neural network}
\newacronym{ula}{ULA}{uniform linear array}

\usepackage{cleveref}

\tikzset{algorithm1/.style={mark options={solid},color=TUMBeamerBlue, line width=\lineWidth, mark=square, dashed}}

\pgfplotscreateplotcyclelist{miko}{%
{color=TUMBeamerRed, dash pattern=on 5pt off 1pt on 1pt off 1pt, line width=1.5pt},%
{color=TUMBeamerOrange, dash pattern=on 5pt off 2pt, line width=1.5pt,
mark=o},%
{color=black, dash pattern=on 5pt off 2pt on 3pt off 2pt, line width=1.5pt},%
{color=TUMDarkerBlue, line width=1.5pt},%
{color=black, dotted, line width=1.5pt},%
{color=TUMBeamerGreen, dotted, line width=1.5pt},%
{color=TUMBeamerYellow, dotted, line width=1.5pt},%
{color=TUMBeamerDarkRed, dotted, line width=1.5pt},%
{color=TUMBeamerLightGreen, dotted, line width=1.5pt},%
{color=TUMIvory, dotted, line width=1.5pt},%
{color=TUMLightBlue, dotted, line width=1.5pt},%
}

\DeclareMathOperator*{\argmax}{arg\,max}
\DeclareMathOperator*{\argmin}{arg\,min}

\DeclareMathOperator{\expec}{E}

\DeclareMathOperator{\vect}{vec}

\newcommand{\calN}{\mathcal{N}}
\newcommand{\calO}{\mathcal{O}}

\newcommand*{\C}{\mathbb{C}}

\newcommand{\herm}{{\operatorname{H}}}
\newcommand{\tp}{{\operatorname{T}}}

\definecolor{myblue}{RGB}{30, 100, 200}

\newlength{\leftstackrelawd}
\newlength{\leftstackrelbwd}
\def\leftstackrel#1#2{\settowidth{\leftstackrelawd}%
	{${{}^{#1}}$}\settowidth{\leftstackrelbwd}{$#2$}%
	\addtolength{\leftstackrelawd}{-\leftstackrelbwd}%
	\leavevmode\ifthenelse{\lengthtest{\leftstackrelawd>0pt}}%
	{\kern-.5\leftstackrelawd}{}\mathrel{\mathop{#2}\limits^{#1}}}

\newcommand{\mbC}{\bm{C}}

\newcommand{\mbH}{\bm{H}}

\newcommand{\mbN}{\bm{N}}

\newcommand{\mbP}{\bm{P}}

\newcommand{\mbU}{\bm{U}}

\newcommand{\mbY}{\bm{Y}}

\newcommand{\mba}{\bm{a}}

\newcommand{\mbh}{\bm{h}}

\newcommand{\mbn}{\bm{n}}

\newcommand{\mby}{\bm{y}}

\newcommand{\mbdelta}{{\bm{\delta}}}

\newcommand{\mbSigma}{{\bm{\Sigma}}}

\newcommand{\mbmu}{{\bm{\mu}}}

\newcommand{\mbzero}{{\bm{0}}}

\newcommand{\hhat}{\hat{\mbh}}

\newcommand{\covhi}{\mbC_i}
\newcommand{\covhk}{\mbC_k}

\usetikzlibrary{intersections,pgfplots.fillbetween,patterns,spy}

\Crefname{figure}{Fig.}{Figs.}
\pgfplotsset{compat=1.15}

\ifdefined\ONECOLUMN

\else

\fi

\definecolor{ourdarkblue}{RGB}{30, 100, 200}
\definecolor{ourdarkgreen}{RGB}{0, 100, 0}
\definecolor{ourdarkorange}{RGB}{201, 98, 18}
\definecolor{ouryellow}{RGB}{220, 210, 50}
\definecolor{myred}{RGB}{255,69,0}
\definecolor{mylila}{RGB}{153,50,204}
\newcommand{\lineWidth}{1.0pt}
\newcommand{\marksize}{2.0pt}
\tikzset{gmmPgmm/.style={mark options={solid},color=TUMBeamerBlue, line width=\lineWidth, mark=square, mark size=\marksize}}
\tikzset{gmmPrnd/.style={mark options={solid},color=TUMBeamerGreen, line width=\lineWidth, mark=o, mark size=\marksize}}
\tikzset{gmmPdft/.style={mark options={solid},color=TUMBeamerOrange, line width=\lineWidth, mark=triangle, mark size=\marksize}}

\tikzset{glmmsePgenie/.style={mark options={solid},color=black, line width=\lineWidth, mark=x, mark size=\marksize}}

\tikzset{dnnPdnn/.style={mark options={solid},color=mylila, line width=\lineWidth, mark=star, mark size=\marksize, dash dot}}

\tikzset{ompPrnd/.style={mark options={solid},color=myred, line width=\lineWidth, mark=diamond, mark size=\marksize, dashed}}
\tikzset{ompPdft/.style={mark options={solid},color=gray, line width=\lineWidth, mark=pentagon, mark size=\marksize, dashed}}

\tikzset{lmmsePrnd/.style={mark options={solid},color=ouryellow, line width=\lineWidth, mark=asterisk, mark size=\marksize, dotted}}
\tikzset{lmmsePdft/.style={mark options={solid},color=brown, line width=\lineWidth, mark=Mercedes star, mark size=\marksize, dotted}}

\newacronym{AWGN}{AWGN}{additive white Gaussian noise}
\newacronym{BLMMSE}{BLMMSE}{Bussgang LMMSE}
\newacronym{BS}{BS}{base station}
\newacronym{CDF}{CDF}{cumulative distribution function}
\newacronym{CNN}{CNN}{convolutional neural network}
\newacronym{CSI}{CSI}{channel state information}
\newacronym{CSIT}{CSIT}{channel state information at the transmitter}
\newacronym{DFT}{DFT}{discrete Fourier transform}
\newacronym{DL}{DL}{downlink}
\newacronym{DNN}{DNN}{deep neural network}
\newacronym{DoA}{DoA}{direction of arrival}
\newacronym{EM}{EM}{expectation maximization}
\newacronym{FDD}{FDD}{frequency division duplex}
\newacronym{GMM}{GMM}{Gaussian mixture model}
\newacronym{LMMSE}{LMMSE}{linear minimum mean square error}
\newacronym{LOS}{LOS}{line of sight}
\newacronym{LS}{LS}{least squares}
\newacronym{MSE}{MSE}{mean squared error}
\newacronym{MIMO}{MIMO}{multiple-input multiple-output}
\newacronym{MPC}{MPC}{multi-path component}
\newacronym{MT}{MT}{mobile terminal}
\newacronym{NLOS}{NLOS}{non-line of sight}
\newacronym{NN}{NN}{neural network}
\newacronym{O2I}{O2I}{outdoor-to-indoor}
\newacronym{OMP}{OMP}{orthogonal matching pursuit}
\newacronym{PDF}{PDF}{probability density function}
\newacronym{PGA}{PGA}{projected gradient ascent}
\newacronym{PSD}{PSD}{power spectral density}
\newacronym{SNR}{SNR}{signal-to-noise ratio}
\newacronym{TDD}{TDD}{time division duplex}
\newacronym{UL}{UL}{uplink}
\newacronym{ULA}{ULA}{uniform linear array}
\newacronym{URA}{URA}{uniform rectangular array}
\newacronym{UMa}{UMa}{urban macrocell}
\newacronym{nSE}{nSE}{normalized spectral efficiency}
\newacronym{cCDF}{cCDF}{complementary cumulative distribution function}
\newacronym{MU-MIMO}{MU-MIMO}{multi-user MIMO}
\newacronym{MU-MISO}{MU-MISO}{multi-user MISO}
\newacronym{BD}{BD}{block diagonalization}
\newacronym{RBD}{RBD}{regularized block diagonalization}
\newacronym{RCI}{RCI}{regularized channel inversion}
\newacronym{WMMSE}{WMMSE}{weighted minimum mean square error}
\newacronym{SWMMSE}{SWMMSE}{stochastic WMMSE}
\newacronym{SVD}{SVD}{singular value decomposition}
\newacronym{SR}{SR}{sum-rate}
\newacronym{CME}{CME}{conditional mean estimator}
\newacronym{ML}{ML}{machine learning}
\newacronym{FLOPS}{FLOPS}{floating-point operations}
\newacronym{OFDM}{OFDM}{orthogonal frequency-division multiplexing}
\newacronym{LTE}{LTE}{Long Term Evolution}
\newacronym{GPS}{GPS}{Global Positioning System}
\newacronym{UMi}{UMi}{urban microcell}
\newacronym{MAP}{MAP}{maximum a posteriori}
\newacronym{3GPP}{3GPP}{3rd Generation Partnership Project}
\newacronym{MMSE}{MMSE}{minimum mean square error}
\newacronym{NMSE}{NMSE}{normalized mean squared error}
\newacronym{MISO}{MISO}{multiple-input single-output}

\newcommand{\Nrx}{N_{\mathrm{rx}}}
\newcommand{\Ntx}{N_{\mathrm{tx}}}

\newcommand{\Krx}{K_{\mathrm{rx}}}
\newcommand{\Ktx}{K_{\mathrm{tx}}}

\begin{document}

\title{Channel-Adaptive Pilot Design for FDD-MIMO Systems Utilizing Gaussian Mixture Models}

\author{Nurettin~Turan, Benedikt~Fesl, Benedikt Böck, Michael Joham, and Wolfgang~Utschick\\
\IEEEauthorblockA{TUM School of Computation, Information and Technology, Technical University of Munich, Germany\\
	Email: \{nurettin.turan, benedikt.fesl, benedikt.boeck, joham, utschick\}@tum.de
    }

\thanks{The authors acknowledge the financial support by the Federal Ministry of
Education and Research of Germany in the program of ``Souver\"an. Digital.
Vernetzt.''. Joint project 6G-life, project identification number: 16KISK002. 
}
\thanks{\copyright This work has been submitted to the IEEE for possible publication. Copyright may be transferred without notice, after which this version may no longer be accessible.
}
}

\maketitle

\begin{abstract}

In this work, we propose to utilize \acp{GMM} to design pilots for \ac{DL} channel estimation in \ac{FDD} systems.
The \ac{GMM} captures prior information during training that is leveraged to design a codebook of pilot matrices in an initial offline phase.
Once shared with the \ac{MT}, the \ac{GMM} is utilized to determine a feedback index at the \ac{MT} in the online phase. 
This index selects a pilot matrix from a codebook, eliminating the need for online pilot optimization.
The \ac{GMM} is further used for \ac{DL} channel estimation at the \ac{MT} via observation-dependent \ac{LMMSE} filters, parametrized by the \ac{GMM}.
The analytic representation of the \ac{GMM} allows adaptation to any \ac{SNR} level and pilot configuration without re-training.
With extensive simulations, we demonstrate the superior performance of the proposed \ac{GMM}-based pilot scheme compared to state-of-the-art approaches.

\end{abstract}


\section{Introduction}

In \ac{MIMO} communication systems, obtaining \ac{CSI} at the \ac{BS} needs to occur in regular time intervals.
In \ac{FDD} systems, both the \ac{BS} and the \ac{MT} transmit at the same time but on different frequencies, which breaks the reciprocity between the instantaneous \ac{UL} \ac{CSI} and \ac{DL} \ac{CSI}.
Consequently, acquiring accurate \ac{DL} \ac{CSI} in \ac{FDD} systems is challenging \cite{2019massive} and thus relies on feedback of the estimated channel from the \ac{MT}.
Therefore, the quality of \ac{DL} channel estimation is of crucial importance.

In massive \ac{MIMO} systems, where the \ac{BS} is typically equipped with a high number of antennas, as many pilots as transmit antennas are required to be sent from the \ac{BS} to the \ac{MT} to fully illuminate the channel, i.e., avoiding a systematic error when relying on \ac{LS} \ac{DL} \ac{CSI} estimation at the \ac{MT}.
However, the associated pilot overhead for complete channel illumination can be prohibitive \cite{BjLaMa16}.
In scenarios with spatial correlation at the \ac{BS} and the \ac{MT}, the \ac{DL} training overhead can be significantly reduced by leveraging statistical knowledge of the channel and the noise \cite{ChLoBi14, KoSa04, BjOt10, PaLiZhLu07, FaLiLiGa17, GuZh19}, e.g., by using Bayesian estimation approaches.

Hence, for \ac{DL} channel estimation, given a budget of $n_\mathrm{p}$ pilots, a common approach to use for channel illumination is transmitting $n_\mathrm{p}$ pilots equivalent to the $n_\mathrm{p}$ dominant eigenvectors of the transmit-side spatial correlation matrix.
However, the aforementioned works rely on either perfect or estimated statistical knowledge at the \ac{BS} and/or at the \ac{MT} side, which may be difficult to acquire.

\emph{Contributions:} In this work, we propose to utilize \acp{GMM} for pilot design.
The proposed scheme neither requires \emph{a priori} knowledge of the channel's statistics at the \ac{BS} nor the \ac{MT}. 
The statistical prior information captured with the \ac{GMM} in the offline phase is exploited to determine a feedback index at the \ac{MT} in the online phase utilizing the \ac{GMM}, which is shared between the \ac{BS} and the \ac{MT}.
This feedback index is sufficient to establish common knowledge of the pilot matrix, which is selected from a codebook of pilot matrices.
Thus, no online pilot optimization is required.
The inference of the feedback information involves computing the responsibilities via the \ac{GMM} and conducting a \ac{MAP} estimation, which selects the index of the \ac{GMM} component with the largest responsibility as the feedback index.
The responsibilities determine how well each component of the \ac{GMM} describes the underlying channel, measured in terms of posterior probabilities.
The responsibilities are additionally used to obtain an estimated channel at the \ac{MT} by performing a convex combination of observation-dependent \ac{LMMSE} filters, which are parametrized by the \ac{GMM} and applied to the observation.
Moreover, the analytic representation of the \ac{GMM} generally allows the adaption to any \ac{SNR} level and pilot configuration without re-training.
Based on extensive simulations, we highlight the superior performance of the proposed \ac{GMM}-based pilot scheme compared to state-of-the-art approaches.

\section{System and Channel Model}
\label{sec:system_channel_model}

We consider a \ac{BS} equipped with $\Ntx$ antennas and a \ac{MT} equipped with $\Nrx$ antennas.
We assume a block-fading model, cf., e.g., \cite{NeWiUt18}, where the \ac{DL} signal for block $t$ received  at the \ac{MT} is expressed as
\begin{equation} \label{eq:noisy_obs}
    \mbY_t = \mbH_t \mbP^\tp_t + \mbN_t \in \C^{\Nrx \times n_\mathrm{p}}
\end{equation}
where $t=0,\dots, T$, with the \ac{MIMO} channel $\mbH_t \in \C^{\Nrx \times \Ntx}$, the pilot matrix $\mbP_t  \in \C^{n_\mathrm{p} \times \Ntx}$, and the \ac{AWGN} $\mbN_t = [\mbn^{\prime}_{t,1}, \dots, \mbn^{\prime}_{t,n_\mathrm{p}}] \in \C^{\Nrx \times n_\mathrm{p}}$ with $\mbn^\prime_{t,p} \sim \mathcal{N}_\C(\bm{0}, \sigma_n^2 \mathbf{I}_{\Nrx})$ for $p \in \{1,2, \dots, n_\mathrm{p}\}$ and $n_\mathrm{p}$ is the number of pilots.
We consider systems with reduced pilot overhead, i.e., $n_\mathrm{p} < \Ntx$.
For the subsequent analysis, it is advantageous to vectorize~\eqref{eq:noisy_obs}:
\begin{equation} \label{eq:noisy_obs_vec}
    \mby_t = (\mbP_t \otimes \mathbf{I}_{\Nrx}) \mbh_t + \mbn_t
\end{equation}
where \( \mbh_t = \vect(\mbH_t) \), \( \mby_t = \vect(\mbY_t) \), \( \mbn_t = \vect(\mbN_t) \), and $\mbn_t \sim \mathcal{N}_\C(\mathbf{0}, \mbSigma)$ with $\mbSigma = \sigma_n^2 \mathbf{I}_{\Nrx n_\mathrm{p}}$.

We adopt the \ac{3GPP} spatial channel model~(see \cite{3GPP_mimo,NeWiUt18}) where channels are modeled conditionally Gaussian, i.e., \( \mbh_t | \mbdelta \sim \calN_\C(\mbzero, \mbC_{\mbdelta}) \).
The covariance matrix $ \mbC_{\mbdelta}$ is assumed to remain constant over $T+1$ blocks.
The random vector \( \mbdelta \) comprises the main angles of arrival/departure of the multi-path propagation cluster between the \ac{BS} and the \ac{MT}.
The main angles of arrival/departure are drawn independently and are uniformly distributed over \( [\frac{-\pi}{2}, \frac{\pi}{2}] \).
The \ac{BS} as well as the \ac{MT} employ a \ac{ULA} such that the transmit- and receive-side spatial channel covariance matrices are given by
\begin{equation}
    \mbC_{\mbdelta}^{\mathrm{\{rx,tx\}}} = \int_{-\pi}^\pi g^{\mathrm{\{rx,tx\}}}(\theta; \mbdelta) \mba^{\mathrm{\{rx,tx\}}}(\theta) \mba^{\mathrm{\{rx,tx\}}}(\theta)^\herm d \theta,
\end{equation}
where $\mba^{\mathrm{\{rx,tx\}}}(\theta) = [1, e^{j\pi\sin(\theta)}, \dots, e^{j\pi(N_{\mathrm{\{rx,tx\}}}-1)\sin(\theta)}]^\tp$ is the array steering vector for an angle of arrival/departure \( \theta \) and \( g^{\mathrm{\{rx,tx\}}}\) is a Laplacian power density whose standard deviation describes the angular spread $\sigma_\text{AS}^\mathrm{\{rx,tx\}}$ of the propagation cluster at the \ac{BS} ($\sigma_\text{AS}^\mathrm{tx}=\SI{2}{\degree}$) and \ac{MT} ($\sigma_\text{AS}^\mathrm{rx}=\SI{35}{\degree}$) side~\cite{3GPP_mimo}.
The overall channel covariance matrix is constructed as \( \mbC_{\mbdelta} = \mbC_{\mbdelta}^{\mathrm{tx}} \otimes \mbC_{\mbdelta}^{\mathrm{rx}} \) due to the assumption of independent scattering in the vicinity of transmitter and receiver, see, e.g., \cite{KeScPeMoFr02}.
In the case of a \ac{MT} equipped with a single antenna, \( \mbC_{\mbdelta} \) degenerates to the transmit-side covariance matrix \( \mbC_{\mbdelta}^{\mathrm{tx}} \).

With
\begin{equation} \label{eq:H_dataset}
     \mathcal{H} = \{ \mbh^\ell \}_{\ell=1}^{L},
\end{equation}
we denote the training data set consisting of $L$ channel samples.
For every channel sample $\mbh^\ell$, we generate random angles, collected in \( \mbdelta^\ell \), and then draw the sample as \( \mbh^\ell \sim \calN_\C(\mbzero, \mbC_{\mbdelta^\ell}) \).
These channels represent a communications environment with unknown \ac{PDF} $f_{\mbh}$.

\section{Pilot Optimization with Perfect Statistical Knowledge} \label{sec:pilot_opt_stat}

Given the knowledge of $\mbdelta$, the observation $\mby_t$ is jointly Gaussian with the channel $\mbh_t$ [see \eqref{eq:noisy_obs_vec}], and we can compute a genie \ac{LMMSE} channel estimate with~\cite{NeWiUt18}
\begin{align}\label{eq:genie_lmmse}
    &\hhat_{t, \text{gLMMSE}} = \expec[\mbh_t \mid \mby_t, \mbdelta] \\
    &= \mbC_{\mbdelta} (\mbP_t \otimes \mathbf{I}_{\Nrx})^\herm ((\mbP_t \otimes \mathbf{I}_{\Nrx}) \mbC_{\mbdelta} (\mbP_t \otimes \mathbf{I}_{\Nrx})^\herm + \mbSigma)^{-1} \mby_t.
\end{align}

The goal of pilot optimization is to design the pilot matrix $\mbP_t$ such that the \ac{MSE} between $\hhat_{t, \text{gLMMSE}}$ and the actual channel $\mbh_t$ is minimized~\cite{FaLiLiGa17, PaLiZhLu07, ChLoBi14}:
\begin{equation}
    \mbP_t^\star = \argmin_{\mbP_t} \ \expec[\|\hhat_{t, \text{gLMMSE}} - \mbh_t\|_2^2]
\end{equation}
where the pilot matrix $\mbP_t$ typically satisfies either a total power constraint as in \cite{FaLiLiGa17, PaLiZhLu07} or an equal power per pilot vector constraint as in \cite{ChLoBi14}.
In this work, we will consider the latter case.
For a given $\mbdelta$, the optimal pilot matrix  $\mbP_t^\star$ for every block is the same, i.e., $\mbP^\star_t = \mbP_\text{genie}$ for all $t=0,\dots, T$.
In particular, $\mbP_\text{genie}$ is a sub-unitary matrix~\cite{ChLoBi14}
\begin{equation} \label{eq:p_genie}
    \mbP_\text{genie} = \sqrt{\rho} \mbU_\mbdelta^\herm[:n_\mathrm{p},:],
\end{equation}
which is composed of the $n_\mathrm{p}$ dominant eigenvectors of the transmit-side covariance matrix \( \mbC_{\mbdelta}^{\mathrm{tx}} = \mbU_\mbdelta \bm{\Lambda}_\mbdelta \mbU_\mbdelta^\herm \) corresponding to the $n_\mathrm{p}$ largest eigenvalues, where $\rho$ denotes the transmit power per pilot vector.
Note that power loading across pilot vectors generally achieves better performance but requires additional processing.
Additionally, with a sub-unitary pilot design, our proposed scheme yields a codebook consisting of pilot matrices that do not depend on the \ac{SNR}, resolving the burden of saving \ac{SNR} level-specific pilot codebooks, see Subsection \ref{sec:pilot_design_gmm}.
\section{GMM-based Pilot Design and Downlink Channel Estimation} \label{sec:gmm_based_scheme}

Any channel $\mbh_t$ of a \ac{MT} located anywhere within the \ac{BS}'s coverage area can be interpreted as a realization of a random variable with \ac{PDF} $f_{\mbh}$ for which however, no analytical expression is available.
Therefore, we utilize a \ac{GMM} to approximate the \ac{PDF} $f_{\mbh}$, similar to \cite{KoFeTuUt21J, TuFeKoJoUt23}.
This learned model is then shared between the \ac{BS} and the \ac{MT} to establish common awareness of the channel characteristics.
The \ac{GMM} is then utilized to infer feedback information for pilot matrix design and for \ac{DL} channel estimation at the \ac{MT} in the online phase. 
Thereby, the feedback information of the \ac{MT} of a preceding fading block $t-1$ is leveraged at the \ac{BS} to select the pilot matrix for the subsequent fading block $t>0$.

\subsection{Modeling the Channel Characteristics at the BS -- Offline} \label{sec:modeling_gmm}

The channel characteristics are captured offline using a \ac{GMM} comprised of $K=2^B$ components,
\begin{equation}\label{eq:gmm_of_h}
    f^{(K)}_{\mbh}(\mbh_t) = \sum\nolimits_{k=1}^K \pi_k \calN_{\C}(\mbh_t; \mbmu_k, \mbC_k)
\end{equation}
where each component of the \ac{GMM} is defined by the mixing coefficient $\pi_k$, the mean $\mbmu_k$, and the covariance matrix $\mbC_k$.

Motivated by the observation that the channel exhibits an unconditioned zero mean and similar to \cite{FeTuBoUt23}, we enforce the means of the \ac{GMM}-components to zero, i.e., $\mbmu_k=\bm{0}$ for all $k \in \{1,\dots, K\}$.
This also reduces the number of learnable parameters and, thus, prevents overfitting.
Note that the parameters of the \ac{GMM}, i.e., $\{\pi_k, \mbC_k\}_{k=1}^K$, remain constant across all blocks.
To obtain maximum likelihood estimates of the \ac{GMM} parameters, we utilize the training dataset \(\mathcal{H} \) [see \eqref{eq:H_dataset}] and employ an \ac{EM} algorithm, as described in~\cite[Subsec.~9.2.2]{bookBi06}, where we enforce the means to zero in every M-step of the \ac{EM} algorithm.

For \ac{MIMO} channels, we further impose a Kronecker factorization on the covariances of the \ac{GMM}, i.e., \( \covhk = \mbC^{\mathrm{tx}}_k \otimes \mbC^{\mathrm{rx}}_k \).
Thus, instead of fitting an unconstrained \ac{GMM} with $N\times N$-dimensional covariances (where $N=\Ntx\Nrx$), we fit separate \acp{GMM} for the transmit and receive sides. 
These transmit-side and receive-side \acp{GMM} possess $\Ntx \times \Ntx$ and $\Nrx \times \Nrx$-dimensional covariances, respectively, with $\Ktx$ and $\Krx$ components. 
Then, by computing the Kronecker products of the corresponding transmit-side and receive-side covariance matrices, we obtain a \ac{GMM} with \( K = \Ktx\Krx \) components and $N\times N$-dimensional covariances.
Imposing this constraint on the \ac{GMM} covariances not only significantly decreases the duration of offline training, facilitates parallelization of the fitting process, and demands fewer training samples due to the reduced number of parameters to be learned, cf. \cite{TuFeKoJoUt23, KoFeTuUt21J}, but also ensures access to a transmit-side covariance during pilot design in the online phase, as discussed in Subsection \ref{sec:pilot_design_gmm}.

Using a \ac{GMM}, we can calculate the posterior probability that the channel $\mbh_t$ originates from component $k$ as~\cite[Sec.~9.2]{bookBi06},
\begin{equation}\label{eq:responsibilities_h}
    p(k \mid \mbh_t) = \frac{\pi_k \calN_{\C}(\mbh_t; \bm{0}, \mbC_k)}{\sum_{i=1}^K \pi_i \calN_{\C}(\mbh_t; \bm{0}, \mbC_i) }.
\end{equation}
These posterior probabilities are commonly referred to as responsibilities.

\subsection{Sharing the Model with the MTs -- Offline}

For a \ac{MT} to infer the feedback information, it must have access to the parameters of the \ac{GMM}.
Conceptually, this involves sharing the model parameters, i.e., $\{\pi_k, \mbC_k\}_{k=1}^K$, with the \acp{MT} upon entering the coverage area of the \ac{BS}.
This transfer is required only once since the \ac{GMM} remains unchanged for a specific \ac{BS} environment.

Incorporating model-based insights to restrict the \ac{GMM} covariances, as discussed in Subsection \ref{sec:modeling_gmm}, additionally significantly reduces the model transfer overhead.
Due to specific antenna array geometries, the \ac{GMM} covariances can be further constrained to a Toeplitz or block-Toeplitz matrix with Toeplitz blocks, in case of a \ac{ULA} or \ac{URA}, respectively, cf.~\cite{TuFeUt23}, with even fewer parameters.
In \cite{TuFeUt23}, it is also further discussed how \acp{GMM} with variable bit lengths can be obtained.
These further structural constraints, as well as the analysis with variable bit lengths, are out of the scope of this work.

\subsection{Inferring the Feedback Information and Estimating the Channel at the MTs -- Online}

In the online phase, the \ac{MT} infers feedback information given the observation $\mby_t$ utilizing the \ac{GMM}.
The joint Gaussian nature of each \ac{GMM} component [see \eqref{eq:gmm_of_h}] combined with the \ac{AWGN}, allows for simple computation of the \ac{GMM} of the observations with the \ac{GMM} from \eqref{eq:gmm_of_h} as
\begin{equation}\label{eq:gmm_y}
\small
    f_{\mby}^{(K)}(\mby_t) = \sum\nolimits_{k=1}^K \pi_k \calN_{\C}(\mby_t; \bm{0}, (\mbP_t \otimes \mathbf{I}_{\Nrx}) \covhk (\mbP_t \otimes \mathbf{I}_{\Nrx})^\herm + \mbSigma).
\end{equation}

Thus, the \ac{MT} can compute the responsibilities based on the observations $\mby_t$ as 
\begin{equation}\label{eq:responsibilities}
\small
    p(k \mid \mby_t) = \frac{\pi_k \calN_{\C}(\mby_t;\bm{0}, (\mbP_t \otimes \mathbf{I}_{\Nrx}) \covhk (\mbP_t \otimes \mathbf{I}_{\Nrx})^\herm + \mbSigma)}{\sum_{i=1}^K \pi_i \calN_{\C}(\mby_t; \bm{0}, (\mbP_t \otimes \mathbf{I}_{\Nrx}) \covhi (\mbP_t \otimes \mathbf{I}_{\Nrx})^\herm + \mbSigma) }.
\end{equation}
The feedback information $k_t^\star$ is then determined through a \ac{MAP} estimation as
\begin{equation} \label{eq:ecsi_index_j}
    k^\star_t = \argmax_{k } ~{p(k \mid \mby_t)}
\end{equation}
where the index of the component with the highest responsibility for the observation $\mby_t$ serves as the corresponding feedback information.
The responsibilities measure how well each component of the \ac{GMM} explains the underlying channel $\mbh_t$ of the observed pilot signal $\mby_t$.
Hence, the feedback information is simply the index of the \ac{GMM} component that best explains the underlying channel.
Subsequently, the responsibilities are utilized to obtain a channel estimate via the \ac{GMM} by calculating a convex combination of per-component \ac{LMMSE} estimates, as discussed in \cite{TuFeKoJoUt23, KoFeTuUt21J}.
In particular, the \ac{MT} estimates the channel by computing
\begin{equation}\label{eq:gmm_estimator_closed_form}
    \hhat_{t,\text{GMM}}(\mby) = \sum_{k=1}^K p(k \mid \mby_t) \hhat_{t,\text{LMMSE},k}(\mby_t),
\end{equation}
using the responsibilities $p(k \mid \mby_t)$ from \eqref{eq:responsibilities} and
\begin{align}\label{eq:lmmse_formula}
\small
    &\hhat_{t,\text{LMMSE},k}(\mby_t) \nonumber\\
    &= \covhk (\mbP_t \otimes \mathbf{I}_{\Nrx})^\herm ((\mbP_t \otimes \mathbf{I}_{\Nrx}) \covhk (\mbP_t \otimes \mathbf{I}_{\Nrx})^\herm + \mbSigma)^{-1}\mby_t .
\end{align}

\subsection{Designing the Pilots at the \ac{BS}  -- Online} \label{sec:pilot_design_gmm}

Consider the eigenvalue decomposition of each of the \ac{GMM}'s transmit-side covariances, i.e., \( \mbC_{k}^{\mathrm{tx}} = \mbU_{k} \bm{\Lambda}_{k} \mbU^\herm_{k} \).
For $t>0$, given the feedback information $k_{t-1}^\star$ of the \ac{MT} from the preceding block $t-1$, we propose employing the pilot matrix $\mbP_t$ at the \ac{BS} for the subsequent block $t$ as [cf. \eqref{eq:p_genie}]
\begin{equation}
    \mbP_t = \sqrt{\rho} \mbU^\herm_{k^\star_{t-1}}[:n_\mathrm{p},:],
\end{equation}
i.e., the $n_\mathrm{p}$ dominant eigenvectors of the $k^\star_{t-1}$-th transmit-side covariance matrix \( \mbC_{k^\star_{t-1}}^{\mathrm{tx}} \) are selected as the pilot matrix.
Since the \ac{GMM}-covariances remain fixed, we can store a set of pilot matrices $\mathcal{P} = \{\mbU^\herm_{k}[:n_\mathrm{p},:]\}_{k=1}^K$, and the online pilot design utilizing the proposed \ac{GMM}-based scheme simplifies to a simple selection task based on the feedback information $k^\star_{t-1}$ from the previous block.
For the initial block $t=0$, we employ a \ac{DFT}-based pilot matrix.

\subsection{Complexity Analysis} \label{sec:comp_ana}

The online computational complexity of the proposed \ac{GMM}-based scheme can be divided into three parts: the inference of the feedback information, the channel estimation, and the pilot design.

Matrix-vector multiplications dominate the computational complexity for inferring the feedback information.
This is because the computation of the responsibilities in~\eqref{eq:responsibilities} entails evaluating Gaussian densities, and the calculations involving determinants and inverses can be pre-computed for a specific \ac{SNR} level due to the fixed \ac{GMM} parameters.
Thus, the inference of the feedback information using \eqref{eq:ecsi_index_j} in the online phase at the \ac{MT} has a complexity of \( \calO(K \Nrx^2 n_\mathrm{p}^2) \).

The additional processing to calculate an estimated channel with the \ac{GMM} via \eqref{eq:gmm_estimator_closed_form} further involves the application of the per-component \ac{LMMSE} filters [see \eqref{eq:lmmse_formula}] which exhibits a computational complexity of \(\calO(K\Nrx^2\Ntx n_\mathrm{p})\), since also the \ac{LMMSE} filters for a given \ac{SNR} level can be pre-computed.
For the feedback inference and application of the \ac{LMMSE} filters, parallelization concerning the number of components $K$ is possible.

The computational complexity of the online pilot design is $\calO(K)$, as it only involves traversing the pre-computed set of pilot matrices $\mathcal{P}$.
Thus, in the online phase, our scheme avoids computing an eigenvalue decomposition, which is required for solving the optimization problem from \eqref{eq:p_genie}.

\section{Baseline Channel Estimators and Pilot Schemes}

\begin{figure}[tb]
    \centering
    \includegraphics[scale=1.0]{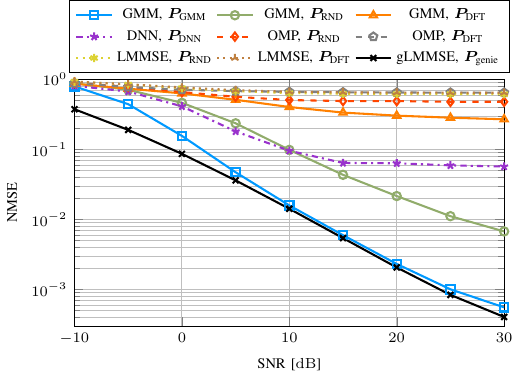}
    \caption{The NMSE over the \ac{SNR} for a MIMO system ($\Ntx=16$, $\Nrx=4$) with $B=7$ feedback bits and $n_\mathrm{p}=4$ pilots.}
    \label{fig:oversnr_16x4_np4_B7}
\end{figure}

In addition to the utopian genie \ac{LMMSE} approach \eqref{eq:genie_lmmse}, where we assume perfect knowledge of $\mbdelta$ (at the \ac{BS} to design the optimal pilots and at the \ac{MT} to apply the genie \ac{LMMSE} estimator), we consider following channel estimators and pilot matrices.

Firstly, we consider the \ac{LMMSE} estimator $\hhat_{\text{LMMSE}}$, where the sample covariance matrix is formed using the set $\mathcal{H}$ [see \eqref{eq:H_dataset}], as discussed in, e.g., \cite{TuFeKoJoUt23, KoFeTuUt21J}.

As another baseline, we consider a compressive sensing estimation method $\hhat_{\text{OMP}}$ employing the \ac{OMP} algorithm, cf.~\cite{TuFeKoJoUt23, AlLeHe15}.

Additionally, we compare to an end-to-end \ac{DNN} approach for \ac{DL} channel estimation with a jointly learned pilot matrix $\mbP_\text{DNN}$, similar to \cite{MaGü21}.
To determine the hyperparameters of the \ac{DNN}, we utilize random search \cite{BeBe12}, with the \ac{MSE} serving as the loss function.
The \ac{DNN} architecture comprises $D_\text{CM}$ convolutional modules, each consisting of a convolutional layer, batch normalization, and an activation function, where $D_\text{CM}$ is randomly selected within $[3, 9]$.
Each convolutional layer contains $D_\text{K}$ kernels, where $D_\text{K}$ is randomly selected within $[32, 64]$.
The activation function in each convolutional module is the same and is randomly selected from $ \{\text{ReLu, sigmoid, PReLU, Leaky ReLU, tanh, swish}\}$.
Following a subsequent two-dimensional max-pooling, the features are flattened, and a fully connected layer is employed with an output dimension of~$2\Ntx\Nrx$ (concatenated real and imaginary parts of the estimated channel).
We train a distinct \ac{DNN} for each pilot configuration and \ac{SNR} level, running $50$ random searches per pilot configuration and \ac{SNR} level and selecting the best-performing \ac{DNN} for each setup.

Lastly, as further baseline pilot matrices, we utilize a \ac{DFT} sub-matrix $\mbP_{\text{DFT}}$ as the pilot matrix, see, e.g., \cite{TsZhWa18}, and alternatively, we consider random pilot matrices denoted by $\mbP_{\text{RND}}$, see, e.g.,~\cite{FaLiLiGa17}.

\section{Simulation Results} \label{sec:sim_results}

\begin{figure}[tb]
    \centering
    \includegraphics[scale=1.0]{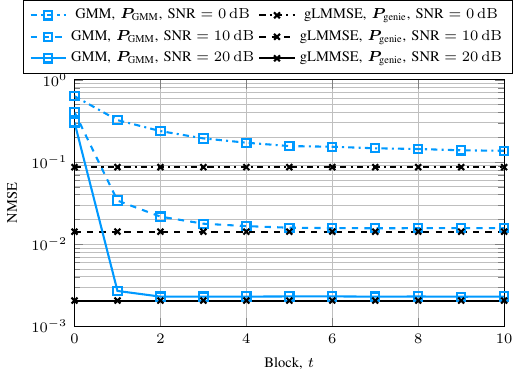}
    \caption{The NMSE over the block index $t$ for a MIMO system ($\Ntx=16$, $\Nrx=4$) with $B=7$ feedback bits, $n_\mathrm{p}=4$ pilots, and $\text{SNR} \in \{\SI{0}{dB}, \SI{10}{dB}, \SI{20}{dB}\}$.}
    \label{fig:overblocks_16x4_np4_B7}
\end{figure}

We use the set $\mathcal{H}$ [see \eqref{eq:H_dataset}] with $L=10^5$ samples for fitting the \ac{GMM} and all other data-based baselines.
We use a different dataset of $J=10^4$ channel samples per block $t$ for evaluation purposes, where we set $T=10$.
The data samples are normalized to satisfy \( \expec[\|\mbh\|^2] = N = \Ntx \Nrx \). 
Additionally, we fix $\rho=1$, enabling the definition of the \ac{SNR} as \( \frac{1}{\sigma_n^2} \).
We employ the \ac{NMSE} as the performance measure.
Specifically, for every block $t$, we compute a corresponding channel estimate \( \hhat^j \)  for each test channel in the set \( \{ \mbh^j \}_{j=1}^J \), and calculate \( \text{NMSE} = \frac{1}{NJ} \sum_{j=1}^{J} \| \mbh^j - \hhat^j \|^2. \)
If not mentioned otherwise, we consider the block with index $t=5$ in the subsequent simulations.

In \Cref{fig:oversnr_16x4_np4_B7}, we simulate a system with $\Ntx=16$ \ac{BS} antennas, $\Nrx=4$ \ac{MT} antennas, and $n_\mathrm{p}=4$ pilots. 
Since we have a \ac{MIMO} setup, we fit a Kronecker structured \ac{GMM} with in total $K=2^7=128$ components ($B=7$ feedback bits), where $\Ktx = 32$ and $\Krx =4$.
The proposed \ac{GMM}-based pilot design scheme denoted by ``GMM, $\mbP_\text{GMM}$'' outperforms all baselines ``\{GMM, OMP, LMMSE\}, \{$\mbP_\text{DFT}$, $\mbP_\text{RND}$\}'' by a large margin, where the \ac{GMM} estimator, the \ac{OMP}-based estimator, or the \ac{LMMSE} estimator, are used in combination with either \ac{DFT}-based pilot matrices or random pilot matrices.
The proposed scheme also outperforms the \ac{DNN} based approach denoted by ``DNN, $\mbP_\text{DNN}$,'' which jointly learns the estimator and a global pilot matrix for the whole scenario; thus, it cannot provide an \ac{MT} adaptive pilot matrix.
This highlights the advantage of the proposed model-based technique over the end-to-end learning technique, which is even trained for each \ac{SNR} level and pilot configuration.
Furthermore, we can observe that the \ac{GMM}-based pilot design scheme performs only slightly worse than the baseline with perfect statistical information at the \ac{BS} and \ac{MT} [see \eqref{eq:genie_lmmse} and \eqref{eq:p_genie}], denoted by  ``gLMMSE, $\mbP_\text{genie}$'', being a utopian estimation approach.
We can observe a larger gap in the low \ac{SNR} regime, where the feedback information obtained through the responsibilities given an observation (see \eqref{eq:ecsi_index_j}) is less accurate due to high noise.

In \Cref{fig:overblocks_16x4_np4_B7}, we analyze the performance of the \ac{GMM}-based pilot design scheme depending on the number of blocks $t$ for the same setup as before at three different \ac{SNR} levels, i.e., $\text{SNR} \in \{\SI{0}{dB}, \SI{10}{dB}, \SI{20}{dB}\}$.
As discussed in Subsection \ref{sec:pilot_design_gmm}, at $t=0$, we utilize \ac{DFT}-based pilots.
After only one block, we can see a significant gain in performance of the proposed \ac{GMM}-based pilot design scheme due to the feedback of the index.
The results further reveal that with an increasing \ac{SNR}, fewer blocks are required to achieve a performance close to the utopian baseline ``gLMMSE, $\mbP_\text{genie}$'' which requires perfect statistical knowledge at the \ac{BS} and the \ac{MT}.

\begin{figure}[tb]
    \centering
    \includegraphics[scale=1.0]{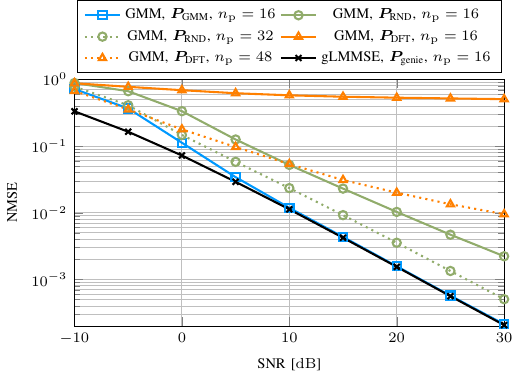}
    \caption{The NMSE over the \ac{SNR} for a MISO system ($\Ntx=64$, $\Nrx=1$) with $B=6$ feedback bits and $n_\mathrm{p} \in \{16,32,48\}$ pilots.}
    \label{fig:oversnr_64_npX}
\end{figure}

In the remainder, we consider a \ac{MISO} system with $\Ntx=64$ \ac{BS} antennas and $\Nrx=1$ antenna at the \ac{MT}.
In \Cref{fig:oversnr_64_npX} we utilize a \ac{GMM} with $K=2^6=64$ components ($B=6$ feedback bits) and consider setups with $n_\mathrm{p} \in \{16,32,48\}$ pilots.
In this case, the proposed scheme ``GMM, $\mbP_\text{GMM}$, $n_\mathrm{p}=16$'', performs only slightly worse than the genie-aided approach ``gLMMSE, $\mbP_\text{genie}$''. 
Moreover, the \ac{GMM}-based pilot design scheme outperforms the baselines ``GMM, \{$\mbP_\text{DFT}$, $\mbP_\text{RND}$\}, $n_\mathrm{p} \in \{16,32,48\}$''.
In particular, the proposed scheme with only $n_\mathrm{p}=16$ pilots outperforms random pilot matrices with twice as much pilots ($n_\mathrm{p}=32$) or in case of \ac{DFT}-based pilots even thrice as much pilots ($n_\mathrm{p}=48$).

\begin{figure}[tb]
    \centering
    \includegraphics[scale=1.0]{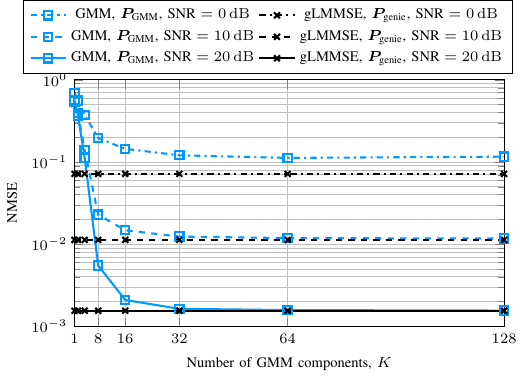}
    \caption{The NMSE over the number of components $K=2^B$ for a MISO system ($\Ntx=64$, $\Nrx=1$) with $n_\mathrm{p} = 16$ pilots, and $\text{SNR} \in \{\SI{0}{dB}, \SI{10}{dB}, \SI{20}{dB}\}$.}
    \label{fig:overcomps_64_np16}
\end{figure}

Lastly, in \Cref{fig:overcomps_64_np16}, we analyze the effect of varying the number of \ac{GMM}-components $K$, on the scheme's performance, where we set $n_\mathrm{p}=16$ and consider three different \ac{SNR} levels, i.e., $\text{SNR} \in \{\SI{0}{dB}, \SI{10}{dB}, \SI{20}{dB}\}$.
We can observe that the estimation error decreases with an increasing number of components $K$.
Moreover, with an increasing \ac{SNR}, the gap to the genie-aided approach ``gLMMSE, $\mbP_\text{genie}$'' decreases.
Above $K=32$ components, a saturation can be observed.
Overall, these results suggest that by varying the number of \ac{GMM} components $K$, a performance-to-complexity trade-off can be realized without sacrificing too much in performance for $K\geq16$ components.

\section{Conclusion}

In this work, we proposed to utilize \acp{GMM} for pilot design in \ac{FDD}-\ac{MIMO} systems.
A significant advantage of the proposed scheme is that it does not require \emph{a priori} knowledge of the channel's statistics at the \ac{BS} and the \ac{MT}.
Instead, it relies on a feedback mechanism, establishing common knowledge of the pilot matrix.
The same \ac{GMM} can be utilized for \ac{DL} channel estimation and can be generally adapted to any desired \ac{SNR} level and pilot configuration without requiring re-training.
Simulation results show that the performance gains achieved with the proposed scheme allow the deployment of system setups with reduced pilot overhead while maintaining a similar estimation performance.
In our future work, we will investigate the extension of the proposed \ac{GMM}-based pilot scheme to multi-user systems, see, e.g., \cite{FaLiLiGa17, GuZh19, JiMoCaNi15, BaXu17}.

\balance
\bibliographystyle{IEEEtran}
\bibliography{IEEEabrv,biblio}
\end{document}